\documentclass[10pt,conference,compsocconf]{IEEEtran}
\usepackage{tikz}
\usepackage{amsmath}
\usepackage{amssymb}
\usepackage{amsthm}
\usepackage{bm}
\usepackage{tcolorbox}
\tcbuselibrary{skins}
\usepackage{lipsum}
\usepackage[linesnumbered]{algorithm2e}
\makeatletter
\renewcommand{\@algocf@capt@plain}{above}
\makeatother
\usetikzlibrary{decorations.pathreplacing}
\usetikzlibrary{arrows}
\usetikzlibrary{snakes}
\usetikzlibrary{calc}
\let\emptyset\varnothing
\definecolor{hotpink}{rgb}{0.9,0,0.5}
\definecolor{deepgray}{gray}{0.35}
\definecolor{deepgray2}{gray}{0.25}
\definecolor{deepgray3}{gray}{0.10}
\definecolor{lightgray}{gray}{0.95}
\definecolor{lightgray2}{gray}{0.85}
\definecolor{purple1}{RGB}{239,229,244}
\definecolor{purple2}{RGB}{216,191,216}
\definecolor{lightblue}{rgb}{0.73,0.33,0.83}
\definecolor{lightpurple}{rgb}{.8,.2,.8}
\definecolor{textcolor}{rgb}{0,0,5}
\definecolor{blue1}{RGB}{187,217,238}
\definecolor{blue2}{RGB}{235,244,250}
\definecolor{yellow1}{RGB}{255,255,102}
\definecolor{blue3}{RGB}{63,40,96}
\definecolor{red1}{RGB}{255, 102, 102}
\definecolor{green1}{RGB}{102, 255, 102}
\newtheorem{lemma}{Lemma}
\newtheorem{thm}{Theorem}
\newtheorem{defn}{Definition}
\theoremstyle{definition}

\def\qedsymbol{\vbox{\hrule\hbox{%
                     \vrule height1.3ex\hskip0.8ex\vrule}\hrule}}
\def\endproof{\qquad\qedsymbol\medskip\par}

\usepackage{cite}

\ifCLASSINFOpdf
\else
\fi

\usepackage{url}
\begin{document}
%
\title{The Right Way to Search Evolving Graphs}


\author{
\IEEEauthorblockN{Jiahao Chen}
\IEEEauthorblockA{Computer Science and Artificial Intelligence Laboratory,\\
Massachusetts Institute of Technology,\\
Cambridge, Massachusetts, 02139-4307, USA\\
jiahao@mit.edu}
\and
\IEEEauthorblockN{Weijian Zhang}
\IEEEauthorblockA{School of Mathematics,\\
University of Manchester,\\
Manchester, M13 9PL, England, UK\\
weijian.zhang@manchester.ac.uk}
}

\maketitle

\begin{abstract}
Evolving graphs arise in problems where interrelations between data change over
time. We present a breadth first search (BFS) algorithm for evolving graphs
that computes the most direct influences between nodes at two different times.
Using simple examples, we show that na\"ive unfoldings of adjacency matrices
miscount the number of temporal paths.
By mapping an evolving graph to an adjacency matrix of an equivalent static graph,
we prove that our generalization of the BFS algorithm correctly accounts for
paths that traverse both space and time.
Finally, we demonstrate how the BFS over evolving graphs can be applied to mine
citation networks.

\end{abstract}

\begin{IEEEkeywords}
~Evolving graph; complex network; breadth first search; data mining.
\end{IEEEkeywords}

%
\IEEEpeerreviewmaketitle

\section{Introduction}

Let's imagine a game played by three people, numbered $1$, $2$, and $3$,
each of whom has a message, labeled $a$, $b$, and $c$ respectively.
At each turn, one particular player is allowed to talk to one other player,
who must in turn convey all the messages in his or her possession.
The goal of the game is to collect all the messages.
Suppose $1$ talks to $2$ first, and $2$ in turn talks to $3$.
Then, $3$ can collect all the messages even without talking to $1$ directly.
However, if $2$ talks to $3$ before $1$ talks to $2$, then
$3$ can never get $a$.

We can analyze the spread of information between the players using graph
theory. In this process, the time ordering of events matters, and hence its
graph representation $G(t) = (V(t), E(t))$ must be time dependent.
Such a graph is called an ``evolving graph''~\cite{fkp89,bkmu12},
``evolving network''~\cite{bfgm08} or ``temporal graph''~\cite{tmml09}.

Treatments of evolving graphs vary in their generality and focus.
Kivel\"a \textit{et al.}~\cite{kabg14} treat time dependence as a special
case of families of graphs with multiple interrelationships.
Others like Flajolet \textit{et al.}~\cite{fkp89} use time to index the family
of related graphs, but are not concerned with explicit time-dependent processes.
Yet others focus on incremental updates to large graphs~\cite{bkmu12}.
Here, we describe evolving graphs as a time-ordered sequence of graphs, similar
to the study of metrized graphs by Tang and coworkers~\cite{ntmm13,tmml09,tsmm09,tmml10} and of
community dynamics by Grindrod, Higham and coworkers~\cite{gphe11,grihig13}.

The game described above can be encoded in an evolving graph.
The spread of information to the winner can be described in terms of traversing
this graph using discrete paths that step in both space and time.
Traversals of an ordinary (static) graph may be computed using well known methods
such as the breadth-first search (BFS). An informal description of BFS generalized
to evolving graphs can be found in \cite{tmml09}.
However, it turns out that na\"ive extensions can lead to incorrect descriptions of
the resulting graph traversals by accounting for traversals of edges in space,
but not necessarily in time. A proper treatment requires the notions of
\emph{node activeness} to describe the set of paths that can only traverse time
or edges, which we call \textit{temporal paths}, as well as \emph{causal edges}
which connect active nodes with the same identity across different times.
As a result, our treatment can be applied to any evolving graph, even those that
are highly dynamic with arbitrary changes to the nodes and edges.

It is well known that sparse matrix-vector product is equivalent to a
one-step BFS on the corresponding (static) graph~\cite[Sec. 1.1]{kegi11}.
In this paper, we demonstrate a correct corresponding result for an evolving
graph by constructing a block triangular matrix representation of the graph that
takes into account both static and causal connections between active nodes.
In Section~\ref{sec:breadth-first-search}, we explain how the BFS algorithm can
be applied to an evolving graph to enumerate paths that traverse edges across time
and space. Section~\ref{sec:temporal-paths} provides an example showing that considering
only products of the time-dependent adjacency matrices fails to enumerate certain
temporal paths. We present and demonstrate the BFS algorithm over evolving graphs
in Section~\ref{sec:evolving-graph-bfs}, showing its formal equivalence to
BFS over a particular static graph generated by adding causal edges that connect
active nodes. This static graph generates an algebraic
representation of the BFS as power iteration of its adjacency matrix to a starting
search node, as shown in Section~\ref{sec:bfsla}. The algebraic formulation also
demonstrates interesting connections between properties of the BFS algorithm and
the adjacency matrix. We describe in Section~\ref{sec:implementation-julia} an
implementation of the algorithm in Julia. Finally in Section~\ref{sec:applications},
we explain how BFS on evolving graphs may be applied to study dynamical processes
over citation networks.

\section{Breadth-First Search over Evolving Graphs}
\label{sec:breadth-first-search}

\subsection{Temporal Paths over Active Nodes}
\label{sec:temporal-paths}

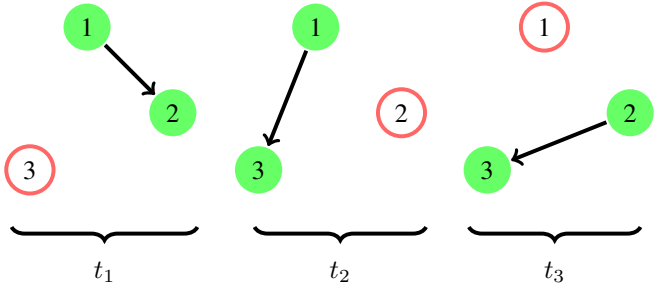
\begin{figure}[h]
 \begin{center}
    \begin{tikzpicture}[scale=.38, line width =1.5pt]
  \node[circle,fill=green1, minimum size=0.2cm] (n7) at (-5,7) {2};
  \node[circle,draw=red1, minimum size=0.2cm] (n8) at (-10,5) {3};
  \node[circle,fill=green1, minimum size=0.2cm] (n10) at (-8,10) {1};

  \node[circle,draw=red1, minimum size=0.2cm] (n6) at (3,7) {2};
  \node[circle,fill=green1, minimum size=0.2cm] (n4) at (-2,5) {3};
  \node[circle,fill=green1, minimum size=0.2cm] (n1) at (0,10) {1};

   \node[circle,fill=green1, minimum size=0.2cm] (n11) at (11,7) {2};
  \node[circle,fill=green1, minimum size=0.2cm] (n12) at (6,5)  {3};
  \node[circle,draw=red1, minimum size=0.2cm] (n14) at (8,10) {1};

  \foreach \from/\to in {n10/n7, n1/n4, n11/n12}
   \draw[every edge,->] (\from) -- (\to);
\draw [decorate,decoration={brace,amplitude=5pt},xshift=-4pt,yshift=0pt]
(4,3) -- (-2,3) node [midway,yshift=-0.6cm]{ $t_2$};
\draw [decorate,decoration={brace,amplitude=5pt},xshift=-4pt,yshift=0pt]
(-4,3) -- (-10.5,3) node [midway,yshift=-0.6cm]{ $t_1$};
\draw [decorate,decoration={brace,amplitude=5pt},xshift=-4pt,yshift=0pt]
(11.5,3) -- (5.5,3) node [midway,yshift=-0.6cm]{ $t_3$};
    \end{tikzpicture}
\end{center}
\caption{An evolving directed graph with 3 time stamps $t_1$, $t_2$ and $t_3$.
At each time stamp, the evolving graph is represented as a graph.
The green filled circles represent active nodes while the red circles represent
inactive nodes. Directed edges are shown as black arrows.}
\label{fig:eg_shortest_path}
\end{figure}

The key new idea in generalizing BFS to evolving graphs is to be able to compute
paths that evolve forward in time and can only traverse the node space along
existing edges. We call these paths \textit{temporal paths}.

Figure~\ref{fig:eg_shortest_path} shows a small example of an evolving
directed graph, $G_3 = \langle G^{[1]}, G^{[2]}, G^{[3]}\rangle$, consisting of
a sequence of three graphs $G^{[i]}$, each bearing a time stamp $t_i$.
There are directed edges $1\rightarrow2$ at time $t_1$,
$1\rightarrow3$ at time $t_2$, and $2\rightarrow3$ at time $t_3$.
Each edge exists only at a particular discrete time and the nodes connected
by edges are considered \textit{active} at that time.

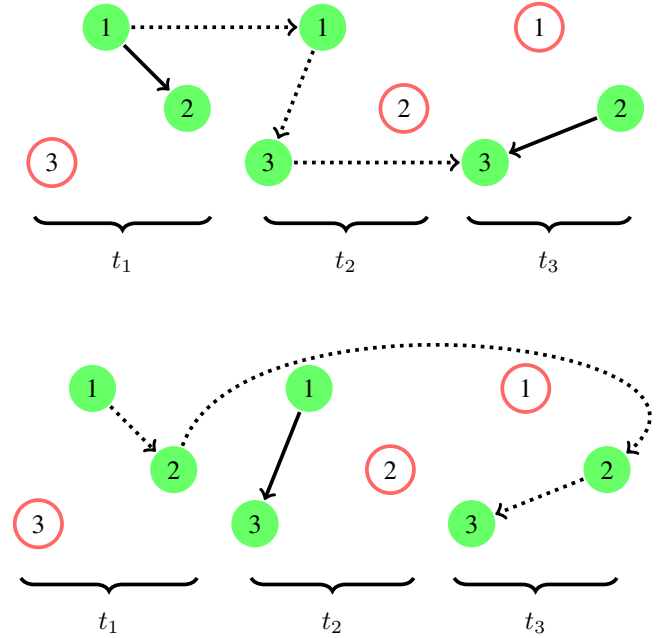
\begin{figure}[h]
 \begin{center}
    \begin{tikzpicture}[scale=.36, line width =1.4pt]
  \node[circle,fill=green1, minimum size=0.2cm] (n7) at (-5,7) {2};
  \node[circle,draw=red1, minimum size=0.2cm] (n8) at (-10,5) {3};
  \node[circle,fill=green1, minimum size=0.2cm] (n10) at (-8,10) {1};

  \node[circle,draw=red1, minimum size=0.2cm] (n6) at (3,7) {2};
  \node[circle,fill=green1, minimum size=0.2cm] (n4) at (-2,5) {3};
  \node[circle,fill=green1, minimum size=0.2cm] (n1) at (0,10) {1};

   \node[circle,fill=green1, minimum size=0.2cm] (n11) at (11,7) {2};
  \node[circle,fill=green1, minimum size=0.2cm] (n12) at (6,5)  {3};
  \node[circle,draw=red1, minimum size=0.2cm] (n14) at (8,10) {1};

  \foreach \from/\to in {n10/n7, n11/n12}
   \draw[every edge,->] (\from) -- (\to);
     \draw[dotted,->](n1) -- (n4);
  \draw[dotted,->](n10) -- (n1);
  \draw[dotted,->](n4) -- (n12);
\draw [decorate,decoration={brace,amplitude=5pt},xshift=-4pt,yshift=0pt]
(4,3) -- (-2,3) node [midway,yshift=-0.6cm]{ $t_2$};
\draw [decorate,decoration={brace,amplitude=5pt},xshift=-4pt,yshift=0pt]
(-4,3) -- (-10.5,3) node [midway,yshift=-0.6cm]{ $t_1$};
\draw [decorate,decoration={brace,amplitude=5pt},xshift=-4pt,yshift=0pt]
(11.5,3) -- (5.5,3) node [midway,yshift=-0.6cm]{ $t_3$};
    \end{tikzpicture}
   \begin{tikzpicture}[scale=.36, line width =1.4pt]
  \node[circle,fill=green1, minimum size=0.2cm] (n7) at (-5,7) {2};
  \node[circle,draw=red1, minimum size=0.2cm] (n8) at (-10,5) {3};
  \node[circle,fill=green1, minimum size=0.2cm] (n10) at (-8,10) {1};

  \node[circle,draw=red1, minimum size=0.2cm] (n6) at (3,7) {2};
  \node[circle,fill=green1, minimum size=0.2cm] (n4) at (-2,5) {3};
  \node[circle,fill=green1, minimum size=0.2cm] (n1) at (0,10) {1};

   \node[circle,fill=green1, minimum size=0.2cm] (n11) at (11,7) {2};
  \node[circle,fill=green1, minimum size=0.2cm] (n12) at (6,5)  {3};
  \node[circle,draw=red1, minimum size=0.2cm] (n14) at (8,10) {1};

  \foreach \from/\to in {n1/n4}
   \draw[every edge,->] (\from) -- (\to);
     \draw[dotted,->](n7) to[out=70, in=40] (n11);
  \draw[dotted,->](n10) -- (n7);
   \draw[dotted,->](n11) -- (n12);
\draw [decorate,decoration={brace,amplitude=5pt},xshift=-4pt,yshift=0pt]
(4,3) -- (-2,3) node [midway,yshift=-0.6cm]{ $t_2$};
\draw [decorate,decoration={brace,amplitude=5pt},xshift=-4pt,yshift=0pt]
(-4,3) -- (-10.5,3) node [midway,yshift=-0.6cm]{ $t_1$};
\draw [decorate,decoration={brace,amplitude=5pt},xshift=-4pt,yshift=0pt]
(11.5,3) -- (5.5,3) node [midway,yshift=-0.6cm]{ $t_3$};
    \end{tikzpicture}
\end{center}
\caption{The two temporal paths of length 4 from $(1, t_1)$ to $(3,t_3)$ on the evolving graph
shown in Figure~\ref{fig:eg_shortest_path}, shown in black dashed lines.
The paths traverse only active nodes along edges, and are allowed to advance
between the same node if it is active at different times.}
\label{fig:active}
\end{figure}

Temporal paths connect only active nodes in ways that respect time ordering.
Thus the sequences
$\langle (1, t_1), (1, t_2), (3, t_2), (3, t_3) \rangle$ and
$\langle (1, t_1), (2, t_1)$, $(2, t_3), (3, t_3) \rangle$
are both examples of \textit{temporal paths} from $(1, t_1)$ to $(3, t_3)$,
which are drawn as dotted lines with arrowheads in Figure~\ref{fig:active}.
However, $\langle (1, t_1), (1, t_2), (2, t_2), (3, t_2), (3, t_3) \rangle$
is not a temporal path because node 2 is inactive at time $t_2$.

The restriction that temporal paths may only traverse active nodes reflects
underlying causal structure in many real world applications, such as analyzing the
influence of nodes over social networks. We will also show later in
Section~\ref{sec:temp-paths-adjac}
that the resulting structure of allowable temporal paths leads to nontrivial
subtleties in the generalization of algorithms and concepts from ordinary
(static) graphs.

\subsection{Breadth-First Traversal Over Temporal Paths}
\label{sec:evolving-graph-bfs}

The example presented above in Section~\ref{sec:temporal-paths}
demonstrates how active nodes restrict the set of temporal paths
that need to be considered when traversing an evolving graph.

We now give a general description of the BFS algorithm over evolving graphs,
both directed and undirected, which correctly takes into account the structure
of temporal paths.
Our notation generalizes that for static graphs presented in \cite{even12,kegi11}.

\begin{defn}
An \textbf{evolving graph} $G_n$ is a sequence of
(static) graphs
$G_n = \langle G^{[1]}, G^{[2]}, \ldots, G^{[n]} \rangle$ with associated
time labels $t_1, t_2, \ldots, t_n$ respectively.
Each $G^{[t]} = (V^{[t]}, E^{[t]})$ represents a (static) graph labeled by a
time $t$.
\end{defn}

Intuitively, an evolving graph is some discretization of the continuous-time family $G(t)$:

\begin{center}
\begin{tikzpicture}[snake=zigzag, line before snake = 5mm, line after snake = 5mm]
    \draw (0,0) -- (2,0);
    \draw[snake] (2,0) -- (4,0);
    \draw (4,0) -- (5,0);
    \draw[snake] (5,0) -- (7,0);

    \foreach \x in {0,1,2,4,5,7}
      \draw (\x cm,3pt) -- (\x cm,-3pt);

    \draw (0,0) node[below=3pt] {$ 1 $} node[above=3pt] {$   $};
    \draw (1,0) node[below=3pt] {$ 2 $} node[above=3pt] {$ G^{[2]} $};
    \draw (2,0) node[below=3pt] {$ 3 $} node[above=3pt] {$  $};
    \draw (3,0) node[below=3pt] {$  $} node[above=3pt] {$  $};
    \draw (4,0) node[below=3pt] {$ 5 $} node[above=3pt] {$  $};
    \draw (5,0) node[below=3pt] {$ 6 $} node[above=3pt] {$ G^{[6]}$};
    \draw (6,0) node[below=3pt] {$  $} node[above=3pt] {$  $};
    \draw (7,0) node[below=3pt] {$ n $} node[above=3pt] {$ $};
  \end{tikzpicture}
\end{center}

We assume no particular relation between the node and edge sets for each static
graph $G^{[t]} = (V^{[t]}, E^{[t]})$. In particular, we allow the node sets to
change over time, so that each $V^{[t]}$ may be different. Changing
node sets happen naturally in citation networks, where nodes may appear or
disappear from the citation network over time.
The addition, removal, or relabeling of nodes can be expressed in terms
of a map $\Pi^{[t,t^\prime]} : V^{[t]} \rightarrow V^{[t^\prime]}$ that expresses
the appropriate permutations and/or projections.

\begin{defn}
A \textbf{temporal node} is a pair $(v, t)$, where $v\in V^{[t]}$ is a node at
a time $t$.
\end{defn}

\begin{defn}
A temporal node $(v,t)$ is an \textbf{active node} if there exists at least one
edge $e\in E^{[t]}$ that connects $v\in V^{[t]}$ to another node $w\in V^{[t]}$,
$w\ne v$.

An \textbf{inactive node} is a temporal node that is not an active node.
\end{defn}

In Figure~\ref{fig:eg_shortest_path}, the temporal nodes $(1,t_1)$ and $(2,t_2)$
are active nodes, whereas the temporal node $(3,t_1)$ is an inactive node.

\begin{defn}
A \textbf{temporal path} of length $m$ on an evolving graph $G_n$
from temporal node $(v_1, t_1)$ to temporal node $(v_m, t_m)$ is
a time-ordered sequence of active nodes,
$\langle (v_1, t_1), (v_2, t_2), \ldots, (v_m, t_m) \rangle$.
Here, time ordering means that $t_1 \leq t_2 \leq \cdots \leq t_m$ and
$v_i = v_j$ iff $t_i \ne t_j$.
\end{defn}

This definition of a temporal path differs from that of the dynamic walk
in \cite{gphe11,grihig13} in that \textbf{causal edges}, i.e.\ sequences of the
form $\langle (v, t), (v, t\prime)\rangle$ are included explicitly in temporal
paths but are only implicitly included in dynamic walks and are not counted
toward the length of dynamic walks.
Our definition implies that if either or both end points of a temporal
path are inactive, then the entire temporal path must be the empty sequence
$\langle \rangle$. Keeping track explicitly of the time labels of each temporal
node allows greater generality to cases where the node sets change over time.
Furthermore, we shall show later in Sec.~\ref{sec:temp-paths-adjac}
that the explicit bookkeeping of the time labels is essential for correctly
generalizing the BFS to evolving graphs.

The following definition of \textbf{forward neighbors} generalizes the notion of
neighbors and reachability in static graphs.

\begin{defn}
The \textbf{$k$-forward neighbors} of a temporal node $(v, t)$ are the temporal nodes
that are the $(k+1)$st temporal node in every temporal path of length $k+1$ starting
from $(v, t)$. The \textbf{forward neighbors} of a temporal node $(v, t)$ are its
1-forward neighbors.
\end{defn}

In Figure~\ref{fig:eg_shortest_path},
the forward neighbors of $(1, t_1)$ are $(2, t_1)$ and $(1, t_2)$ and
the only forward neighbor of $(2, t_1)$ is $(2, t_3)$.
The 2-forward neighbors of $(1, t_1)$ are
$(2, t_1), (1, t_2), (2, t_2)$ and $(3, t_2)$.
By construction, time stamp of every forward neighbor of an active node $(v, t)$
must be no earlier than $t$.

\begin{defn}
The \textbf{distance} from a temporal node $(v, t)$ to a temporal node $(w, s)$ is
the $k$ for which $(w, s)$ is a $k$-forward neighbor of $(v, t)$.
\end{defn}

Our definition of distance, again, differs from the definition of distance
in the formulation of \cite{gphe11,grihig13} in that we explicitly count causal
edges toward the distance. It also differs from the notion of temporal distance
in the work of Tang and coworkers~\cite{tmml10}, which is the number of time
steps between $t$ and $s$ (inclusive). In this respect, our formulation of the
BFS on evolving graphs differs from these other works by minimizing a different
notion of distance over an evolving graph.

Note that this notion of distance is not a metric, since the distance from $(v, t)$
to $(w, s)$ will in general differ from the distance of $(v, t)$ from $(w, s)$
owing to time ordering.

\begin{defn}
A temporal node $(w, s)$ is \textbf{reachable} from a temporal node $(v, t)$ if
the distance to $(w, s)$ from a temporal node $(v, t)$ there exists some finite
integer $k$ for which $(w, s)$ is a $k$-forward neighbor of $(v, t)$.
\end{defn}

\subsection{Description of the BFS algorithm}
\label{sec:descr-algor}

The BFS on evolving graphs is described in Algorithm~\ref{alg:bfs}.
Algorithm~\ref{alg:bfs} is identical to the BFS on static graphs except for
line $8$, where we visit the forward neighbors of a given temporal node
in both space and time.
Given an evolving graph $G_n$ and a root $(v_1, t_1)$,
Algorithm~\ref{alg:bfs} returns all temporal notes reachable from the root and
their distances from the root.
$reached$ is a dictionary from temporal nodes to integers whose key set represents
all visited temporal nodes and whose value set are the corresponding distances
from the root.

The BFS constructs a tree inductively by discovering all $k$-forward neighbors
of the root before proceeding to all $(k+1)$-forward neighbors of the root.
Within the outermost loop, the algorithm iterates over $frontier$, a list of all
temporal nodes of distance $k$ from the root. The $nextfrontier$ list is populated
with all temporal nodes that are forward neighbors of any temporal node in the
$frontier$ list which have not yet been reached by the algorithm.

 \begin{center}
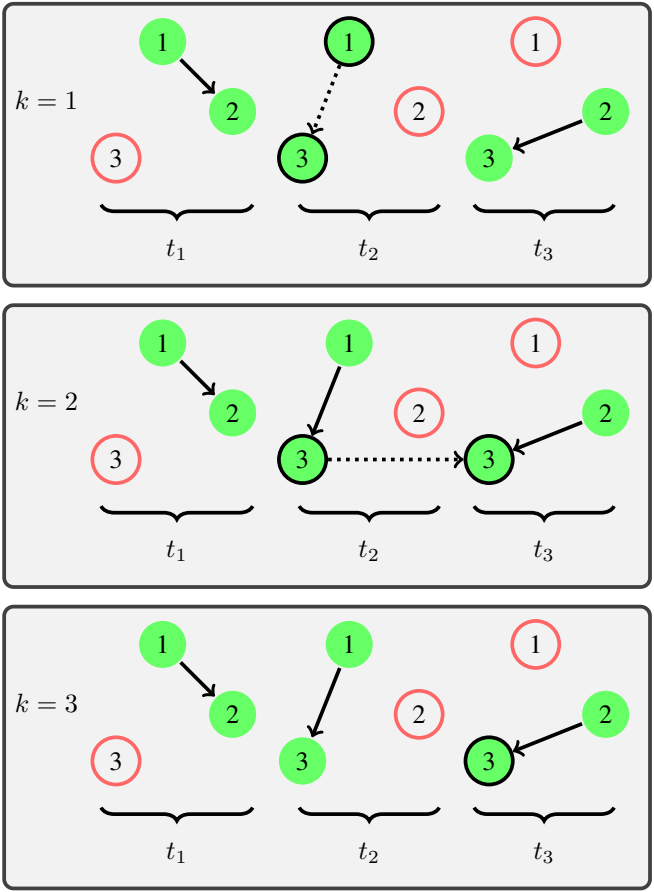
\begin{figure}[h]
\begin{tcolorbox}[width=3.4in,
                  boxsep=0pt,
                  left=0pt,
                  right=0pt,
                  top=4pt,
                  ]
    \begin{tikzpicture}[scale=.31, line width =1.4pt]
 \node[text width=3cm] at (-9.5, 7.5) {$k=1$};
  \node[circle,fill=green1, minimum size=0.2cm] (n7) at (-5,7) {2};
  \node[circle,draw=red1, minimum size=0.2cm] (n8) at (-10,5) {3};
  \node[circle,fill=green1, minimum size=0.2cm] (n10) at (-8,10) {1};

  \node[circle,draw=red1, minimum size=0.2cm] (n6) at (3,7) {2};
  \node[circle,fill=green1, draw=black, minimum size=0.2cm] (n4) at (-2,5) {3};
  \node[circle,fill=green1, draw=black, minimum size=0.2cm] (n1) at (0,10) {1};

   \node[circle,fill=green1, minimum size=0.2cm] (n11) at (11,7) {2};
  \node[circle,fill=green1, minimum size=0.2cm] (n12) at (6,5)  {3};
  \node[circle,draw=red1, minimum size=0.2cm] (n14) at (8,10) {1};

  \foreach \from/\to in {n10/n7, n11/n12}
   \draw[every edge,->] (\from) -- (\to);
     \draw[dotted,->](n1) -- (n4);
\draw [decorate,decoration={brace,amplitude=5pt},xshift=-4pt,yshift=0pt]
(4,3) -- (-2,3) node [midway,yshift=-0.6cm]{ $t_2$};
\draw [decorate,decoration={brace,amplitude=5pt},xshift=-4pt,yshift=0pt]
(-4,3) -- (-10.5,3) node [midway,yshift=-0.6cm]{ $t_1$};
\draw [decorate,decoration={brace,amplitude=5pt},xshift=-4pt,yshift=0pt]
(11.5,3) -- (5.5,3) node [midway,yshift=-0.6cm]{ $t_3$};
    \end{tikzpicture}
\end{tcolorbox}
\begin{tcolorbox}[width=3.4in,
                  interior hidden,
                  boxsep=0pt,
                  left=0pt,
                  right=0pt,
                  top=4pt,
                  ]
   \begin{tikzpicture}[scale=.31, line width =1.4pt]
 \node[text width=3cm] at (-9.5, 7.5) {$k=2$};
  \node[circle,fill=green1, minimum size=0.2cm] (n7) at (-5,7) {2};
  \node[circle,draw=red1, minimum size=0.2cm] (n8) at (-10,5) {3};
  \node[circle,fill=green1, minimum size=0.2cm] (n10) at (-8,10) {1};

  \node[circle,draw=red1, minimum size=0.2cm] (n6) at (3,7) {2};
  \node[circle,fill=green1, draw=black, minimum size=0.2cm] (n4) at (-2,5) {3};
  \node[circle,fill=green1, minimum size=0.2cm] (n1) at (0,10) {1};

   \node[circle,fill=green1, minimum size=0.2cm] (n11) at (11,7) {2};
  \node[circle,fill=green1, draw=black, minimum size=0.2cm] (n12) at (6,5)  {3};
  \node[circle,draw=red1, minimum size=0.2cm] (n14) at (8,10) {1};

  \foreach \from/\to in {n1/n4, n10/n7, n11/n12}
   \draw[every edge,->] (\from) -- (\to);
  \draw[dotted,->](n4) -- (n12);
\draw [decorate,decoration={brace,amplitude=5pt},xshift=-4pt,yshift=0pt]
(4,3) -- (-2,3) node [midway,yshift=-0.6cm]{ $t_2$};
\draw [decorate,decoration={brace,amplitude=5pt},xshift=-4pt,yshift=0pt]
(-4,3) -- (-10.5,3) node [midway,yshift=-0.6cm]{ $t_1$};
\draw [decorate,decoration={brace,amplitude=5pt},xshift=-4pt,yshift=0pt]
(11.5,3) -- (5.5,3) node [midway,yshift=-0.6cm]{ $t_3$};
    \end{tikzpicture}
\end{tcolorbox}
\begin{tcolorbox}[width=3.4in,
                  interior hidden,
                  boxsep=0pt,
                  left=0pt,
                  right=0pt,
                  top=4pt,
                  ]
   \begin{tikzpicture}[scale=.31, line width =1.4pt]
 \node[text width=3cm] at (-9.5, 7.5) {$k=3$};
  \node[circle,fill=green1, minimum size=0.2cm] (n7) at (-5,7) {2};
  \node[circle,draw=red1, minimum size=0.2cm] (n8) at (-10,5) {3};
  \node[circle,fill=green1, minimum size=0.2cm] (n10) at (-8,10) {1};

  \node[circle,draw=red1, minimum size=0.2cm] (n6) at (3,7) {2};
  \node[circle,fill=green1, minimum size=0.2cm] (n4) at (-2,5) {3};
  \node[circle,fill=green1, minimum size=0.2cm] (n1) at (0,10) {1};

   \node[circle,fill=green1, minimum size=0.2cm] (n11) at (11,7) {2};
  \node[circle,fill=green1, draw = black, minimum size=0.2cm] (n12) at (6,5)  {3};
  \node[circle,draw=red1, minimum size=0.2cm] (n14) at (8,10) {1};

  \foreach \from/\to in {n1/n4, n10/n7, n11/n12}
   \draw[every edge,->] (\from) -- (\to);
\draw [decorate,decoration={brace,amplitude=5pt},xshift=-4pt,yshift=0pt]
(4,3) -- (-2,3) node [midway,yshift=-0.6cm]{ $t_2$};
\draw [decorate,decoration={brace,amplitude=5pt},xshift=-4pt,yshift=0pt]
(-4,3) -- (-10.5,3) node [midway,yshift=-0.6cm]{ $t_1$};
\draw [decorate,decoration={brace,amplitude=5pt},xshift=-4pt,yshift=0pt]
(11.5,3) -- (5.5,3) node [midway,yshift=-0.6cm]{ $t_3$};
    \end{tikzpicture}
\end{tcolorbox}
\caption{Breadth-first search (BFS) on the evolving graph shown in
Figure~\ref{fig:eg_shortest_path} starting from the root $(1, t_2)$ at iteration
$k = 1, 2, 3$.
Note that the time $t_1$ does not participate in the BFS. Black circles indicate
the active nodes forming the $frontier$ and $nextfrontier$ sets in Algorithm~\ref{alg:bfs},
connected by the dotted black lines.}
\label{fig:example}
\end{figure}
\end{center}
As a simple example, consider the BFS on the example graph in
Figure~\ref{fig:eg_shortest_path} starting from the root $(1, t_2)$.
The procedure is shown in Figure \ref{fig:example}.
The $frontier$ list is first initialized to $\{(1,t_2)\}$.
Since the only forward neighbor of $(1,t_2)$ is $(3,t_2)$,
iteration $k=1$ produces $reached[(3,t_2)]=1$ and $nextfrontier = \{(3,t_2)\}$.
In the next iteration $k=2$, the only forward neighbor of $(3,t_2)$ is $(3,t_3)$,
so $reached[(3,t_3)]=2$ and $nextfrontier = \{(3,t_3)\}$.
The algorithm terminates after $k=3$ after verifying that $(3,t_3)$ has no
forward neighbors.

The preceding example illustrates the fact that $G^{[1]}$ plays no part in the
BFS traversal of $G_n$ starting from $(1, t_2)$. In general, all $G^{[t]}$ with
time stamps $t < t'$ for a starting node $(v, t')$ are irrelevant to the BFS
traversal. Hence without loss of generality we may assume that BFS is always
computed with a root at time $t_1$, the earliest time stamp in $G_n$.

\begin{algorithm}[h]
 \SetKwFor{For}{for}{}{end}
 \SetKwFor{While}{while}{}{end}
 \SetKwFor{If}{if}{}{end}
 \SetKwFunction{nei}{forwardneighbors}
 \SetKwProg{Fn}{function}{}{end}
 \Fn{BFS($G_n, (v_1,t_1)$)}{
  $reached[(v_1,t_1)] = 0$ \\
  $frontier = \{(v_1,t_1)\}$ \\
  $k = 1$ \\
  \While{$frontier \ne \emptyset$}{
   $nextfrontier = \emptyset$ \\
   \For{$(v,t) \in frontier$}{
     \For{$(v',t') \in$\nei{$(v,t)$}}{
          \If{$(v',t') \notin reached$}{
              $reached[(v',t')] = k$\\
              $nextfrontier = nextfrontier \cup \{(v',t')\}$ \\
              }
         }
       }
  $frontier = nextfrontier$ \\
  $k = k+1$ \\
  }
  \textbf{return} $reached$
}
\caption{Breadth-first search (BFS) on an evolving graph $G_n$ starting from a
root $(v_1, t_1)$. The return value, $reached$, is a dictionary mapping all
reachable temporal notes from the root to their distances from the root. At the
end of each iteration $k$, the $frontier$ set contains all temporal nodes of
distance $k$ from the root.
}
\label{alg:bfs}
\end{algorithm}

\begin{thm}[Correctness of the evolving graph BFS]
Let $G_n$ be an evolving graph and $(v_1, t_1)$ be an active node of $G_n$.
Then Algorithm~\ref{alg:bfs} discovers every active node that is reachable from
the root $(v_1, t_1)$, and $reached[(v, t)]$ is the distance from $(v_1, t_1)$
to $(v, t)$.
\label{thm:correct}
\end{thm}

\IEEEproof
Let's first consider the case when $G_n$ is directed.
Define the set of temporal nodes
$\tilde V_L^{[t]} = \{(v_1, t) | (v_1, v_2) \in E^{[t]}\}$, which
consists of the active nodes at time $t$ which participate on the left side of an edge.
Similarly, $\tilde V_R^{[t]} = \{ (v_2, t) | (v_1, v_2) \in E^{[t]}\}$
contains the corresponding active nodes on the right side of an edge. Then
$\tilde V^{[t]} = \tilde V_L^{[t]} \cup \tilde V_R^{[t]}$ is the set of active
nodes at time $t$, and $V = \bigcup_t \tilde V^{[t]}$ is the set of all
active nodes in $G_n$.

Similarly, define the set of \textbf{causal edges} $E' = \{(u_s,v_t)|u_s=(u,s)\in V,
v_t=(v,t)\in V, v=u, s<t\}$, which consists of temporal nodes that connect
active nodes sharing the same node at different times. Each edge in $E'$ is then
in 1-1 correspondence with a temporal path of length 2,
$\langle (v,s), (v,t) \rangle$.
Define also the set of \textbf{static edges at time $t$},
$\tilde E^{[t]} = \{(e, t)|e\in E^{[t]}\}$, which are simply the edge
sets in $G_n$ with time labels, and the set of \textbf{static edges} $\tilde E$,
being simply the union over all times, $\bigcup_t \tilde E^{[t]}$.
Then $E = \tilde E \cup E'$ is the set of all edges representing all allowed
temporal paths of length 2.

The node set $V$ and edge set $E$ now define a static directed graph $G=(V,E)$
that is in 1-1 correspondence with the evolving graph $G_n$. The node set $V$ of $G$ is in 1-1 correspondence with active nodes of $G_n$ while the edge set $E$ is in 1-1
correspondence with all temporal paths of length 2 on $G_n$.

We now establish a similar 1-1 correspondence of forward neighbors of an active
node with a subset of $G$. By induction, all new nodes populated into the key
set of $reached$ at iteration $k$ are of distance $k$ from the root. By definition,
the forward neighbors of some active node $(v, t) \in G_n$ are
active nodes of either the form $(v, t')$ for some $t' > t$ or $(u, t)$ for some
$u\ne v$. In other words, they are connected either by a causal edge or a static edge.
Clearly, the former are elements of $E' \subseteq E$ while the latter
are elements of $\tilde E \subseteq E$.
Thus each forward neighbor of an active node $(v, t) \in G_n$ is in 1-1
correspondence with a node in $V$ that is a neighbor of $v_t \in V$.

When $G_n$ is undirected, every edge in $\tilde E^{[t]}$ can be represented by
two edges in $G$: from an active node in $\tilde V_L^{[t]}$ to
an active node in $\tilde V_R^{[t]}$ and the reverse.
Every edge in $E'$ is in 1-1 correspondence with
an edge in $G$ by causality. Therefore, the forward neighbors of an
active node is in 1-1 correspondence with a subset of $G$ and the analysis above
follows.

The correctness of BFS on the evolving graph $G_n$ now follows from the
correctness of BFS on the static graph $G$, since we have also established a 1-1
correspondence for every intermediate quantity in Algorithm~\ref{alg:bfs}.
\endproof

As presented, the BFS over evolving graphs makes no assumptions about how the
evolving graph $G_n$ is represented. Suppose it is represented by a collection of
adjacency lists, one for each active node in $G_n$. Then we have that the
asymptotic complexity of BFS on $G_n$ is the same as that for BFS on $G$, using
the 1-1 construction of $G$ from $G_n$.

\begin{thm}[Computational complexity of the evolving graph BFS]
\label{thm:complexity}
Let $G_n$ be an evolving graph represented using adjacency lists,
$(v_1, t_1)$ be an active node of $G_n$, and $G=(V,E)$ be the static graph
constructed from $G_n$ using the 1-1 correspondences defined in the proof of
Theorem~\ref{thm:correct}. Then the asymptotic computational complexity of
Algorithm~\ref{alg:bfs} is $O(|E| + |V|)$.
\end{thm}

\IEEEproof
Any edge in any edge set of $G_n$ can be accessed in constant time in random
access memory. By construction, BFS on $G_n$ is in 1-1 correspondence with BFS
on the static graph $G$. The number of operations of BFS on $G$ is $O(|E| + |V|)$,
and so the result follows.
\endproof

Note that in the theorems in this section construct an equivalent static graph
$G$ corresponding to the evolving graph $G_n$. However, $G$ contains more edges
than the union of all the static parts of $G_n$, as we also add causal edges
$E'$. To our knowledge, our formulation of the BFS represents the first attempt
to include these edges explicitly in the treatment of evolving graphs.

\section{Formulating the Evolving Graph BFS with Linear Algebra}
\label{sec:representation}

\subsection{The importance of causal edges}
\label{sec:temp-paths-adjac}

For each static graph $G^{[t]} = (V^{[t]}, E^{[t]})$ that constitutes the
evolving graph $G_n$, define its corresponding
$\left\vert V^{[t]}\right\vert \times \left\vert V^{[t]}\right\vert$
one-sided adjacency matrix with elements
\begin{equation}
A^{[t]}_{ij} =
\begin{cases}
 1 & \mbox{if $(i,j)\in E^{[t]}$,} \\
 0 & \mbox{otherwise.}
\end{cases}
\label{eqn:adjacency}
\end{equation}

We can then represent $G_n$ using a sequence of adjacency matrices
$A_n = \langle A^{[1]}, A^{[2]}, \ldots, A^{n]}\rangle$. The example in
Figure~\ref{fig:eg_shortest_path} can be represented as
\[
\left\langle
  \begin{bmatrix}
    0 & 1 & 0 \\
    0 & 0 & 0 \\
    0 & 0 & 0
  \end{bmatrix},~
 \begin{bmatrix}
   0 & 0 & 1 \\
   0 & 0 & 0 \\
   0 & 0 & 0
 \end{bmatrix},~
 \begin{bmatrix}
  0 & 0 & 0 \\
  0 & 0 & 1 \\
  0 & 0 & 0
 \end{bmatrix}
\right\rangle.
\]

For a static graph $G$ with adjacency matrix $A$, $(A^k)_{ij}$ counts the number
of paths of length $k$ between node $i$ and node $j$. Na\"ively, one might want
to generalize this result to evolving graphs by postulating that the $(i,j)$th
entry of the discrete path sum
\begin{align}
S^{[t_n]} = A^{[t_1]}A^{[t_n]} + \sum_{t_1 \le t \le t_n} A^{[t_1]}A^{[t]}A^{[t_n]}
+ \cdots \nonumber \\
+ \sum_{t_1 \le t \le t' \le \dots \le t_n}
A^{[t_1]}A^{[t]}A^{[t']}\cdots A^{[t_n]}
\label{eq:sum}
\end{align}
counts the number of temporal paths from $(i,t_1)$ to $(j,t_n)$.
However, this postulate is incorrect.
In the example of Figure~\ref{fig:eg_shortest_path},
\[
(S^{[t_3]})_{13} = \left(A^{[t_1]}A^{[t_2]}A^{[t_3]} + A^{[t_1]}A^{[t_3]}\right)_{13} = 1
\]
even though there are clearly two temporal paths from
$(1, t_1)$ to $(3, t_3)$ as shown in Figure~\ref{fig:active}.

The first term in the sum $S^{[t_3]}$ vanishes since $A^{[t_1]}A^{[t_2]} = 0$.
Furthermore, the vanishing of $S^{[t_2]} = A^{[t_1]}A^{[t_2]}$ itself reflects the absence of any temporal path
from $t_1$ to $t_2$ that goes through at least one edge at $t_1$. However,
\begin{equation}
\langle(1, t_1), (1, t_2), (3, t_2)\rangle~\label{eq:tpathex}
\end{equation}
is a clearly a valid temporal path as shown in Figure~\ref{fig:active} which
cannot be expressed by a product of adjacency matrices.

Sums $S^{[t]}$ of the form \eqref{eq:sum} produce an incorrect count of temporal
paths because they do not capture temporal paths with causal edges, i.e.\ subpaths
of the form $\langle (v, s), (v, t) \rangle$, $s < t$.
One might attempt to amend the sums $S^{[t]}$ in \eqref{eq:sum} by redefining
the adjacency matrices to include ones along the diagonal, hence allowing
paths containing the sequence $\langle (i, t_1), (i, t_2) \rangle$. However, the resulting
sum is still incorrect, as it counts paths with subsequences $\langle (3, t_1), (3, t_2) \rangle$
and are hence not temporal paths. Instead, the temporal path \eqref{eq:tpathex}
is counted by the matrix product $M^{[t_1, t_2]} A^{[t_2]}$, where
\begin{equation}
    M^{[t_1, t_2]} = \begin{pmatrix}1 & 0 & 0 \\ 0 & 0 & 0 \\ 0 & 0 & 0\end{pmatrix}.
\label{eq:m-example}
\end{equation}
$M^{[t_1, t_2]}$ describes the forward time propagation of temporal nodes that
are active at both times $t_1$ and $t_2$, i.e.\ it counts
temporal paths that contain subsequences $\langle (i, t_1), (i, t_2) \rangle$,
and both $(i, t_1)$ and $(i, t_2)$ are active nodes.

The simple example of Figure~\ref{fig:eg_shortest_path} demonstrates why sums
over products of adjacency matrices of the form \eqref{eq:sum} do not count
temporal paths correctly: they neglect the combinatorics associated with
the causal edge set $E'$. In the next section, we show how to account for these
causal edges by introducing a new matrix--vector product $\odot$.


\subsection{Defining forward neighbors algebraically}

The algebraic representation of evolving graphs presented in Section \ref{sec:temp-paths-adjac}
allows us to exploit a graphical interpretation of
matrix--vector products involving the adjacency matrix~\cite{kegi11}.
If $A$ is the adjacency matrix of a (static) graph $G$ and $e_k$ is the $k$th
elementary unit vector, then the nonzero entries of $A^T e_k$ have indices that
are neighbors of $k$.
The algebraic formulation of BFS on evolving graphs follows similarly, but
requires a new kind of matrix--vector product, $\odot$, defined by

\[
A^T \!\odot b =
\begin{cases}
b & \mbox{if $A^Tb \ne 0$ or $Ab \ne 0$,} \\
0 & \mbox{otherwise.}
\end{cases}
\]
The condition $A^Tb \ne 0$ ensures that the product is nonzero in components
involving left active nodes $\cup_t \tilde V_L^{[t]}$, and the condition
$Ab \ne 0$ is the analogue for right active nodes $\cup_t \tilde V_R^{[t]}$.
The forward neighbors of a temporal node $(k, t_1)$ in $A_n$ can then be determined
from the indices and time stamps of the nonzero elements in the sequence
\begin{equation}
\label{eq:out_ne}
\big\langle (A^{[1]})^Te_k, \; (A^{[2]})^T\!\odot e_k, \; \ldots \; (A^{[n]})^T\!\odot e_k\big\rangle.
\end{equation}
The nonzero entries of the first vector represent forward neighbors that are on
the same time stamp $t_1$, whereas nonzero entries of the other vectors represent forward
neighbors that are advanced in time but remain on the same node $k$.
The quantity \eqref{eq:out_ne} therefore encodes a BFS tree of depth 2,
as its nonzero entries are labeled by all temporal nodes of distance 1 from
$(k, t_1)$.

Referring back to the example of Figure~\ref{fig:eg_shortest_path}, the forward
neighbors of node $(1, t_1)$ can be computed by
\begin{align*}
\left\langle
  \begin{bmatrix}
    0 & 0 & 0 \\
    1 & 0 & 0 \\
    0 & 0 & 0
  \end{bmatrix}\!\!\!\!
\begin{bmatrix}
1 \\
0 \\
0
\end{bmatrix}\!,
 \begin{bmatrix}
   0 & 0 & 0 \\
   0 & 0 & 0 \\
   1 & 0 & 0
 \end{bmatrix}\!\!
\odot\!\!
\begin{bmatrix}
1 \\
0 \\
0
\end{bmatrix}\!,
 \begin{bmatrix}
  0 & 0 & 0 \\
  0 & 0 & 0 \\
  0 & 1 & 0
 \end{bmatrix}\!\!
\odot\!\!
\begin{bmatrix}
1 \\
0 \\
0
\end{bmatrix}\!
\right\rangle
\\
=
\left\langle
\begin{bmatrix}
0 \\
1 \\
0
\end{bmatrix},~
\begin{bmatrix}
1 \\
0 \\
0
\end{bmatrix},~
\begin{bmatrix}
0 \\
0 \\
0
\end{bmatrix}
\right\rangle
\end{align*}
From this computation, we can deduce that $(2,t_1)$ and $(1,t_2)$ are the forward
neighbors of $(1,t_1)$.

\subsection{Evolving graphs as a blocked adjacency matrix}

The proof of Theorem~\ref{thm:correct} provides a construction for representing
an evolving graph $G_n$ by a static graph $G$ with nodes corresponding to active
nodes of $G_n$. It turns out that the block structure of $G$ is useful for
understanding the nature of the $\odot$ operation.

Consider the second iteration of BFS on $G_n$ with root $(k, t_1)$, which
requires computing the sequences

\begin{subequations}
\begin{align}
  \label{eq:5}
 \big\langle (A^{[1]})^Tc_1,   (A^{[2]})^T\odot c_1, \ldots,  (A^{[n]})^T\odot c_1  \big\rangle  \\
\big\langle (A^{[2]})^Tc_{2}, \ldots, (A^{[n]})^T\odot c_{2} \big\rangle \\
 \ldots  \\
  \big\langle (A^{[n]})^Tc_n \big\rangle
\label{eq:6}
\end{align}
\end{subequations}
where $c_1 = (A^{[1]})^Te_k$ and $c_i = (A^{[i]})^T \odot e_k$ for $i>1$.
Summing resultant vectors that share the same
time stamp, we obtain vectors whose nonzero elements have indexes labeled by
the forward neighbors of the nodes computed at step $1$.

Compare this with the matrix

\[
\bm M_n = \begin{bmatrix}
A^{[t_1]} & M^{[t_1, t_2]} & ... & M^{[t_1, t_n]} \\
        0 & A^{[t_2]}      & ... & M^{[t_2, t_n]} \\
                           &   ... \\
        0 & 0 & ... & A^{[t_n]}
\end{bmatrix}
\]
where $M^{[t_i, t_j]}$ is the matrix whose rows are labeled by $V^{[t_i]}$ and
columns are labeled by $V^{[t_j]}$, and whose entries are
\[
M^{[t_i, t_j]}_{uv} =
\begin{cases}
1 & \mbox{if $(u,v)\in E^\prime$,} \\
0 & \mbox{otherwise.}
\end{cases}
\]
The adjacency matrix blocks $A^{[t]}$ encode the static edge set $\tilde E$,
whereas the off-diagonal blocks $M^{[t_i, t_j]}$ together encode the causal edge
set $E^\prime$, which capture temporal paths with subsequences of the form
$\langle (v, t_i), (v, t_j) \rangle$. Then $\bm M_n$ is the adjacency matrix of
the graph $(\cup_t V^{[t]}, E)$, which is the graph $G$ together with all the
inactive nodes. From the definition, $\bm M_n$ has nonzero
entries only in rows and columns that correspond to active nodes $V$, and so
retaining only these rows and columns corresponding to $V$ produces the
adjacency matrix $\bm A_n$ of $G = (V, E)$.

The off-diagonal blocks $M^{[t_i, t_j]}$ provide an explicit matrix representation
for the $\odot$ product in that $(M^{[t_i, t_j]})^T b = (A^{[t_i]})^T \odot b$.
An example of such an off-diagonal block was already provided in \eqref{eq:m-example}.
These off-diagonal blocks represent traversal between active notes with the same
node space labels but are still separated by time, and are essential for
the correct enumeration of temporal paths.
The upper triangular structure of $\bm M_n$ (and hence $\bm A_n$) reflects the causal
nature of temporal paths in that they cannot go backward in time.

The BFS algorithm presented above can therefore be interpreted as
computing the sequence of matrix--vector products $\bm b$, $\bm A_n^T \bm b$,
$(\bm A_n^T)^2 \bm b$, ..., formed by applying successive monomials of $\bm A_n^T$ to the
block vector
$\bm b^T = [b^T, 0, \cdots, 0 ]$
where $b^T$ encodes the root in the space of active nodes $\tilde V^{[t_1]}$.

\begin{figure}[h]
 \begin{center}
   \begin{tikzpicture}[scale=.36, line width =1.4pt]
  \node[circle,fill=green1, minimum size=0.1cm, inner sep = 0pt] (n7) at (-5,7)
{\scriptsize $(2,t_1)$};
  \node[circle,draw=red1, minimum size=0.2cm, inner sep = 0pt] (n8) at (-10,5)
{\scriptsize $(3,t_1)$};
  \node[circle,fill=green1, minimum size=0.1cm, inner sep = 0pt] (n10) at (-8,10) {\scriptsize $(1,t_1)$};

  \node[circle,draw=red1, minimum size=0.2cm, inner sep = 0pt] (n6) at (3,7)
{\scriptsize $(2,t_2)$};
  \node[circle,fill=green1, minimum size=0.2cm,  inner sep = 0pt] (n4) at (-2,5)
{\scriptsize $(3,t_2)$};
  \node[circle,fill=green1, minimum size=0.2cm,  inner sep = 0pt] (n1) at (0,10)
{\scriptsize $(1,t_2)$};

   \node[circle,fill=green1, minimum size=0.2cm,  inner sep = 0pt] (n11) at (11,7)
{\scriptsize $(2,t_3)$};
  \node[circle,fill=green1, minimum size=0.2cm,  inner sep = 0pt] (n12) at (6,5)
{\scriptsize $(3,t_3)$};
  \node[circle,draw=red1, minimum size=0.2cm,  inner sep = 0pt] (n14) at (8,10)
{\scriptsize $(1,t_3)$};

  \foreach \from/\to in {n10/n7, n1/n4, n11/n12}
   \draw[every edge,->] (\from) -- (\to);
     \draw[dotted,->](n7) to[out=80, in=40] (n11);
  \draw[dotted,->](n10) to[out=10, in=180] (n1);
   \draw[dotted,->](n4) to[out=-30,in=-150] (n12);
    \end{tikzpicture}
\end{center}
\caption{Static graphs corresponding to the evolving graph example of
Figure~\ref{fig:eg_shortest_path}. The green nodes are active nodes while the
red nodes are inactive nodes.
The black lines are edges in the static edge set $\tilde E$ and are encoded
algebraically in the diagonal blocks $A^{[t]}$ of the adjacency matrix $\bm A_3$
or $\bm M_3$.
The dotted lines are edges in the causal edge set $E'$ and are encoded algebraically in
the off-diagonal blocks $M^{[t_i, t_j]}$.
The static graph $G$ constructed in the proof of Theorem~\ref{thm:correct} is
formed by retaining all the edges shown and only the active nodes, and has the
adjacency matrix $\bm A_3$. The graph containing all the edges and temporal nodes
shown has adjacency matrix $\bm M_3$.}
\label{fig:static}
\end{figure}
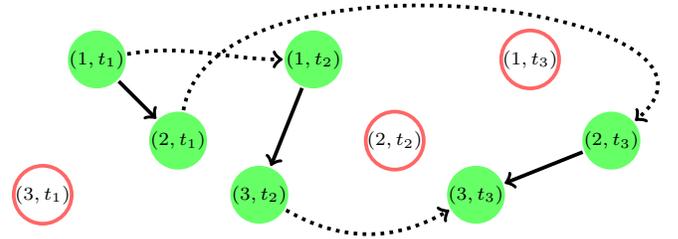

For the example of Figure~\ref{fig:eg_shortest_path}, we have
\begin{align*}
V & = \{(1,\!t_1), (2,\!t_1), (1,\!t_2), (3,\!t_2), (2,\!t_3), (3,\!t_3)\},\\
\tilde E &= \{((1,\!t_1), (2,\!t_1)), ((1,\!t_2), (3,\!t_2)), ((2,\!t_3), (3,\!t_3))\},\\
E'  &= \{((1,\!t_1), (1,\!t_2)), ((2,\!t_2), (2,\!t_3)), ((3,\!t_2), (3,\!t_3))\}.
\end{align*}
In the order specified for $V$, the adjacency matrix of $G$ is then
\[
\bm A_3 = \begin{bmatrix}
0 & 1 & 1 & 0 & 0 & 0 \\
0 & 0 & 0 & 0 & 1 & 0 \\
0 & 0 & 0 & 1 & 0 & 0 \\
0 & 0 & 0 & 0 & 0 & 1 \\
0 & 0 & 0 & 0 & 0 & 1 \\
0 & 0 & 0 & 0 & 0 & 0
\end{bmatrix}
\]

Starting from the vector $\bm b = e_1$, the sequence of iterates is then
\[
\left\langle
\begin{bmatrix}1 \\ 0 \\ 0 \\ 0 \\ 0 \\ 0 \end{bmatrix},
\begin{bmatrix}0 \\ 1 \\ 1 \\ 0 \\ 0 \\ 0 \end{bmatrix},
\begin{bmatrix}0 \\ 0 \\ 0 \\ 1 \\ 1 \\ 0 \end{bmatrix},
\begin{bmatrix}0 \\ 0 \\ 0 \\ 0 \\ 0 \\ 2 \end{bmatrix},
\begin{bmatrix}0 \\ 0 \\ 0 \\ 0 \\ 0 \\ 0 \end{bmatrix},
\dots
\right\rangle.
\]

We see that $(\bm A_n^T)^2$ encodes all the products in
\eqref{eq:5}-\eqref{eq:6}, including both the ordinary matrix--vector product
and the $\odot$ product. Furthermore, $(\bm A_3^T)^3 \bm b$ correctly counts the
two allowed temporal paths from $(1, t_1)$ to $(3, t_3)$, and that the
off-diagonal structure encoded in $E'$ and the $M^{[t,t']}$ blocks are critical
to obtaining the correct count.

Finally, we note that some results regarding the BFS on evolving graphs can be
derived easily using properties of the block adjacency matrix $\bm A_n$. For
example, we can prove a simple lemma that the block adjacency matrix $\bm A_n$ is
nilpotent whenever all the subgraphs $G^{[t]}$ of $G_n$ are acyclic.

\begin{lemma}[Acyclicity implies nilpotence]
Let $G_n = \langle G^{[t_i]} \rangle_{i=1}^n$ be an evolving directed graph and
let all the directed graphs $G^{[t]}$ be acyclic. Then $\bm A_n$ is nilpotent.
\label{thm:nilpotent}
\end{lemma}

\IEEEproof
Recall from the definition of the matrix $\bm A_n$ that it is block upper
triangular, reflecting causality.
Since each directed graph $G^{[t]}$ is acyclic, its corresponding adjacency
matrix $A^{[t]}$ is strictly upper triangular. As a result, $\bm A_n$ must be
upper triangular.

Furthermore, none of the graphs $G^{[t]}$ can have any self-edges,
i.e.\ edges of the form $(u, u)$, and so all diagonal entries of
$A^{[t]}$ must be zero.  Therefore all the diagonal entries of $\bm A_n$ by
construction must be zero also.

We have now proven that $\bm A_n$ is an upper triangular matrix whose diagonal
entries are all zero. Therefore, $\bm A_n$ is nilpotent.
\endproof

Lemma~\ref{thm:nilpotent} also holds for acyclic undirected graphs so long as
the corresponding adjacency matrix representation is encoded in an asymmetric
fashion akin to \eqref{eqn:adjacency}.

The blocked matrix structure of the adjacency matrices presented here provide
interesting relationships between their matrix properties and the algorithmic
properties of BFS, made possible because of the reformulation of BFS as
repeated power iterations of the adjacency matrix in Algorithm~\ref{alg:bfsla}.
Note, however, that these matrices need never be instantiated for practical
computations. Rather, since Algorithm~\ref{alg:bfsla} only requires the
matrix--vector product involving the adjacency matrix, the formulation of
Algorithm~\ref{alg:bfs} provides an efficient way to exploit the block structure
of $\bm A_n$. The $\odot$ operation provides an efficient way to compute the
action of the off-diagonal products. Representing the diagonal blocks
$A^{[t]}$ as sparse matrices further reduces the cost of BFS by exploiting
latent sparsity in graphs that show up in practical applications.

\subsection{The algebraic formulation of BFS on evolving graphs}
\label{sec:bfsla}

The blocked matrix--vector products introduced in the previous section allows us
to write down an elegant algebraic formulation of BFS on evolving graphs, as
presented in Algorithm~\ref{alg:bfsla}.

\begin{algorithm}[h]
 \SetKwFor{For}{for}{}{end}
 \SetKwFor{While}{while}{}{end}
 \SetKwFor{If}{if}{}{end}
 \SetKwFunction{nei}{forward-neighbors}
 \SetKwFunction{enqueue}{enqueue}
 \SetKwFunction{nz}{nonzeros}
 \SetKwFunction{intersect}{intersect}
 \SetKwFunction{map}{activeNodes}
 \SetKwProg{Fn}{function}{}{end}
 \Fn{ABFS($A_n, (v_1,t_1)$)}{
   Form $\bm A^T_n$ from $A_n$.\\
   $\bm{b}_{v_1} = 1$ \\
   $k = 1$ \\
   $reached[(v_1,t_1)] = 0$ \\
   \While{\nz{$\bm{b}$} $\ne \emptyset$}{
       $\bm{b} = \bm{A}^T_n\bm{b}$ \\
       \For{$k \in$ \nz{$\bm{b}$}}{
         \If{\map{$k$} $\in reached$}{
           $\bm{b}_k = 0$ \\
         }
      }
       \For{$node \in $\map{$\bm{b}$}}{
           $reached[node] = k$ \\
         }
       $k = k + 1$
       }
  \textbf{return} $reached$ \\
}
\caption{An algebraic formulation of BFS on evolving graphs.
Given $A_n$, the adjacency matrix representation of $G_n$ and $(v_1,t_1)$, a
node of $G_n$, returns $reached$ as defined in Algorithm~\ref{alg:bfs}.
The function \texttt{nonzeros(v)} returns the nonzero indices of the vector $v$,
and the function \texttt{map(b)} maps a block vector's indices to their corresponding
active nodes.}
\label{alg:bfsla}
\end{algorithm}

\begin{thm}
Algorithm~\ref{alg:bfsla} terminates.
\end{thm}

\IEEEproof
First, we prove that the BFS terminates in the case of acyclic evolving graphs.
Recall from Lemma~\ref{thm:nilpotent} that $\bm A_n$ is nilpotent, i.e.\
there exists some positive integer $k$ for which $\bm A_n^k = 0$.
Hence, after iteration $k$, $\bm b$ is assigned the value $\bm A_n^k \bm b = 0$.
Therefore, Algorithm~\ref{alg:bfsla} must terminate after iteration $k$.

For evolving graphs with cycles, lines 9-11 of Algorithm \ref{alg:bfsla} enforce
that the BFS visits each active node at most once. Since the $k$th block of
$\bm b$ is zeroed out if an active node has already been visited, the subgraph
traversed in the BFS cannot have cycles. Thus all that is required is the
previous result that the BFS on an acyclic graph terminates.
\endproof

\begin{thm}
Algorithm \ref{alg:bfs} and Algorithm \ref{alg:bfsla} are equivalent.
\end{thm}

\IEEEproof
The initialization steps are trivially equivalent.
At the beginning of iteration $k$, the block vector $\bm b$ represents the
frontier nodes encoding the $frontier$ set of Algorithm~\ref{alg:bfs}.
The matrix--vector product $\bm A_n^T \bm b$ encodes the forward neighbors of all
the frontier nodes. Subsequently, active nodes that have already been visited in
previous iterations are zeroed out of the new $\bm b$.
\endproof

\subsection{Computational complexity analysis of the algebraic BFS}
\label{sec:comp-compl-analys}

The complexity of the algebraic BFS of Algorithm \ref{alg:bfsla} is significantly
more complicated that that of Algorithm \ref{alg:bfs}. While the latter uses the
usual adjacency list representation for graphs, the computational cost of the
former depends critically on the actual representation of the matrices.
Furthermore, the average case analysis is complicated by the expected fill-in of
the vector $\bm b$, which influences the cost of the matrix-vector product on line 7
and the expected number of iterations of the \textbf{while} loop beginning on line 6.

While a full complexity analysis is beyond the scope of this paper, it is
straightforward to present worst-case results for dense and compressed sparse
column (CSC) matrices.

\begin{lemma}[Number of iterations]
In the worst case, the number of iterations in the \textbf{while} loop of
Algorithm~\ref{alg:bfsla} is $k = O(|E|)$.
\end{lemma}

\IEEEproof
In the worst case, the BFS must traverse every active node, and only one new
active node is discovered in each iteration. The number of active
nodes is bounded above by the cardinality of the full edge set, $|E|$.
\endproof

The average case analysis for the number of iterations is considerably more
complicated and is beyond the scope of this paper.

\begin{thm}[Dense matrices]
Suppose $\bm A_n$ is represented as a dense matrix. Then the computational
complexity of Algorithm~\ref{alg:bfsla} is $O(k |V|^2)$, which in the worst case is
$O(|E| |V|^2)$.
\end{thm}

\IEEEproof
Since $\bm A_n$ is a $|V| \times |V|$ matrix, the matrix-vector product
$\bm A_n \bm b$ takes $O(|V|^2)$ operations to compute. Thus the cost of
Algorithm~\ref{alg:bfsla} is $O(k |V|^2) = O(|E| |V|^2)$ in the worst case.
\endproof

It is clear that practical implementations of the BFS should never construct
the full matrix $\bm A_n$ in memory. What happens if we use a sparse blocked
representation?

\begin{thm}[Block diagonal sparse matrices]
Suppose $\bm A_n$ is represented by a collection of compressed sparse column
matrices for each diagonal block $A^{[t]}$. Then the
computational complexity of Algorithm~\ref{alg:bfsla} is $O(k(|\tilde E| + |V|))$, which in
the worst case is $O(|E|(|\tilde E| + |V|))$.
\end{thm}

\IEEEproof
The gaxpy operation for CSC matrices costs $2n_{nz}$ flops, where $n_{nz}$ is the
number of stored values in the matrix. The cost of each diagonal subblock calculation
$A^{[t]} b_t$ is therefore $O(|E^{[t]}|)$, since $A^{[t]}$ by construction has
nonzero entries only when a static edge exists.

The off-diagonal products
$M^{[t,t\prime]} b_{t\prime}$ can be computed in $O(|V^{[t]}| + |E^{[t]}|)$ time
for all $t\prime \ge t$
since it can be implemented by the $\cdot$ operation, which constructs either a
zero vector or keeps the same vector.
The cost of checking the condition $(A^{[t]})^T b_{t^\prime} \ne 0$ is
$O(|E^{t}|)$ in the worst case since all that is required is to check whether or
not each column of $A$ is empty.
Similarly, checking the condition $A^{[t]} b_{t^\prime} \ne 0$ reduces to
checking if each row of $A$ is empty, and thus is of cost $O(|V^{t}|)$.

Thus, the cost of multiplying one block row of $\bm A^T$ (for some time $t$)
with $\bm b$ is $O(|V^{[t]}| + |E^{[t]}|)$. Summing over all times yields the
desired result.
\endproof

We can see that even implementing the BFS algebraically using CSC matrices is
insufficient to reduce the running time to linear, which can be achieved for
the adjacency list representation in Algorithm \ref{alg:bfs}.
This result strongly suggests that additional work is needed to produce true
algorithmic equivalence at the computational level.


\section{Implementation in Julia}
\label{sec:implementation-julia}

To study evolving graphs and experiment with various graph types, we have
developed EvolvingGraphs.jl~\cite{zhang15}, a software package for the creation,
manipulation, and study of evolving graphs written in Julia~\cite{bkse12}.
It is freely available online at
\begin{center}
\url{https://github.com/weijianzhang/EvolvingGraphs.jl}
\end{center}
and available with the MIT ``Expat'' license. The package contains an implementation
of the evolving graph BFS of Algorithm~\ref{alg:bfs}. \texttt{IntEvolvingGraph}, a data type in EvolvingGraphs.jl,
represents an evolving graph as adjacency lists.

We now present some simple timing data to show that our implementation of
Algorithm~\ref{alg:bfs} is indeed linear scaling in computational cost.

\begin{figure}[h]
  \centering
  \includegraphics[scale=0.58]{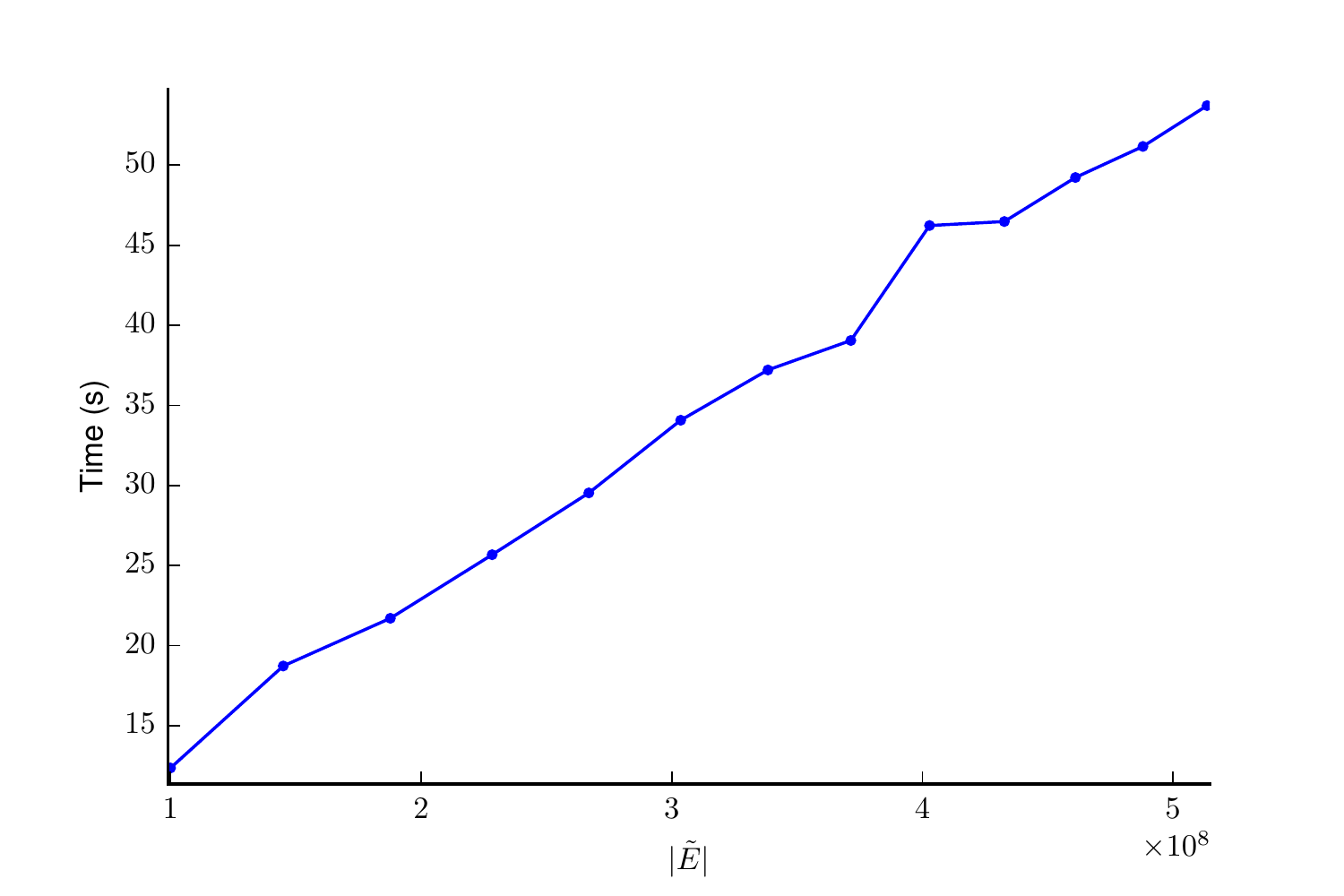}
  \caption{Experimental run time of Algorithm \ref{alg:bfs}
on a collection of random evolving graphs with $10^5$ active nodes and $10$ time
stamps, showing linear scaling in $|\tilde E|$. The horizontal axis shows the
number of edges ($|\tilde E|$, including only the static edges) of each evolving
graph, while the vertical axis shows the corresponding
computation time.}
  \label{fig:time}
\end{figure}

We generate a sequence of random (directed) \texttt{IntEvolvingGraph}s
with $10^5$ active nodes and $10$ time stamps.
The first \texttt{IntEvolvingGraph} in the sequence has about
$10^8$ static edges.
We consecutively add new
random static edges to this \texttt{IntEvolvingGraph}. For example,
the second random \texttt{IntEvolvingGraph} in the sequence
has about $1.5 \times 10^8$ static edges and the third
has $1.8 \times 10^8$ static edges.
Note that in this experiment, we do not have direct control
over the full edge set $E$, only the static edge set $\tilde E$.
When we add new static edges, new causal edges may be added as well
(if the corresponding temporal nodes were not active before).
However, the number of newly introduced causal edges for each active node
is bounded by the number of time stamps, so it suffices to demonstrate
linear scaling in $|\tilde E|$.
Figure \ref{fig:time} shows the plots of number of edges against the computation time
for running Algorithm \ref{alg:bfs} in Julia. All experiments are conducted
on a single core of a Linux system with 1TB of RAM and 80 cores of Intel(R)
Xeon(R) E7-8850s running at 2.00 GHz clock speed.
The results show Algorithm \ref{alg:bfs} can be computed in linear time, which agrees
with the result of Theorem \ref{thm:complexity}.

\section{Application to Citation Networks}
\label{sec:applications}

Evolving graphs have found many applications to analyzing networks that change
over time \cite{gphe11,grihig13}. In this section, we focus specifically on
citation networks, and show that evolving graph formalism presented above can be
used to capture the dynamical structure of citation networks. Consider
the evolving graph $G_n=\langle G^{[t]} \rangle_t$ such that $G^{[t]}$ has
node set corresponding to authors active at time $t$ and directed edge set
$E^{[t]} \ni (i, j)$ representing a citation of author $j$ by author $i$ in a
publication at time $t$.

Then given an author $a$ at time $t_1$, the evolving graph BFS described above
can compute $T(a, t_1)$, the set of all the authors that have been influenced by
$a$'s work at time $t_1$.
Define also a \emph{community} to be a group of
researchers that have been influenced by the same authors.
For example, given a paper published by $a$ at time $t$,
we can determine $a$'s community by searching backward in time to find
$T^{-1}(a, t)$, the authors that influenced $a$ at time $t$, and then searching
forward to find $T(l_1, t_1) \cup T(l_1, t_2) \cup \cdots \cup T(l_k, t_k)$, where
$(l_1,t_1), (l_2, t_2), \ldots, (l_k, t_k)$ are the leaves of $T^{-1}(a,t)$.
The backward search in time follows straightforwardly from the forward time traversal
presented above simply by reversing the time labels, e.g.\ by the transformation
$t\rightarrow -t$.

We are currently investigating the use of our evolving graph BFS on citation
networks.

\section{Conclusion}

The correct generalization of BFS to evolving graphs necessitates a careful
enumeration of temporal paths. The structure associated with causal edges $E'$
turns out to be of vital importance, and cannot be capture simply by products of
successive adjacency matrices, which by construction can only capture the
topologies of the static edges $\tilde E$. Only by considering both causal edges
and static edges can we show that BFS over any evolving graph $G_n$ computes the
correct result for our notion of distance.
The new concepts of activeness, temporal paths, and causal edges make possible a
correct implementation of BFS to evolving graphs and we expect that these
ideas will continue to provide powerful new insights into how similar graphical
algorithms may be generalized correctly.

Furthermore, we show that BFS on evolving graphs admits an algebraic formulation
that easily provides nontrivial results, such as termination of the algorithm.
However, our current understanding tells us that the BFS over evolving graphs
is most efficiently computed in the adjacency list representation, thus never
forming explicit matrix-vector products. Further work is needed to elucidate
more efficient formulations of the algebraic BFS for evolving graphs.


\section*{Acknowledgments}

We thank N. J. Higham and V. \u Sego (Manchester) for helpful suggestions.
W.\ Z.\ thanks the School of Mathematics at U. Manchester for research \& travel funding and
A. Edelman for arranging for a fruitful visit to MIT.



\bibliographystyle{IEEEtran}
%

\bibliography{strings,njhigham,paper}

\end{document}